\documentclass[prl,twocolumn,showpacs]{revtex4}

\usepackage{epsfig,color}
\begin{document}
\newcommand{\ltwid}{\mathrel{\raise.3ex\hbox{$<$\kern-.75em\lower1ex\hbox{$\sim$}}}}
\newcommand{\gtwid}{\mathrel{\raise.3ex\hbox{$>$\kern-.75em\lower1ex\hbox{$\sim$}}}}
\newcommand{\bra}{\langle}
\newcommand{\ket}{\rangle}
\newcommand{\sill}{\psi}
\newcommand{\trace}{{\rm Tr}}
\newcommand{\ntilde}{\tilde{n}}
\newcommand{\stilde}{\tilde{s}}
\newcommand{\atilde}{\tilde{\alpha}}
\newcommand{\israel}[1]{\textcolor{red}{#1}}
\newcommand{\adrian}[1]{\textcolor{blue}{#1}}
\def\nn{\nonumber\\}

\bibliographystyle{apsrev}

\title{Hermitian and non-Hermitian thermal Hamiltonians}

\author{Adrian E. Feiguin}
\affiliation{Department of Physics, Northeastern University, Boston, Massachusetts 02115, USA}

\author{Israel Klich}
\affiliation{Department of Physics, University of Virginia, Chalottesville, VA, USA}


\begin{abstract}
Thermal density matrices can be described by a pure quantum state within the thermofield formalism. Here we show how to construct a class of Hamiltonians realizing a thermofield state as their ground state. These Hamiltonians are frustration-free, and can be Hermitian or non-Hermitian, allowing one to use ground-state methods to understand the thermodynamic properties of the system. In particular this approach gives an explicit mapping of thermal phase transitions into quantum phase transitions. In the non-Hermitian case, the quantum phase transition is not accompanied by a change in the spectrum of the Hamiltonian, which remains gapped. We illustrate these ideas for the classical 2D Ising model.
\end{abstract}
\maketitle

\paragraph{Introduction.} While physical systems at low and high temperatures are usually unrelated and require completely different sets of tools to study, the existence of possible relations between both regimes in various systems is a fascinating topic. Perhaps, the most famous result of this kind is the duality between low and high temperature expansions of spin systems since Krammers and Wannier used it to exactly locate the critical point for the Ising model on a square lattice \cite{Krammers}.

Here we present a simple framework by which high temperature behavior of quantum systems is described using low temperature methods, by finding the ground state of an appropriate Hamiltonian. In particular, for systems whose Hamiltonians consist of local, mutually commuting operators, the mapping can be done exactly at any temperatures. Such models consist, naturally, of classical systems, thereby allowing us to make a direct connection between classical critical points and quantum critical points at the same dimension. For classical models, our approach is complementary to the representation of such states in terms of classical thermal coherent states (see e.g. \cite{henley2004classical,somma2007quantum}). Other problems of interest, such as topological Hamiltonians as the Toric code are also amenable to an exact mapping in this way.


Our first aim is to find a Hamiltonian whose ground state realizes a thermofield dynamics scheme \cite{Takahashi1975,UmezawaBook,Matsumoto1986}. 
The idea of representing a thermal state as a pure state in a larger space has been around for a long time, however, usually its focus is on the properties of the wave function itself. Here, we take the idea in a new direction, namely we ask: what kind of Hamiltonian on the larger system, has this ``purified'' thermal state as its ground state?  


Let us first recall the basic idea of TFD. Suppose that we want to represent an operator $A$
\[
A=\sum_{jk}a_{jk} |j\rangle \langle k|.
\] 
If the dimension of the Hilbert space is $d$, then we need $d\times d$ entries $a_{jk}$ to define $A$ (which are obviously basis dependent). Another generic form to represent an operator is by working in Liouville space, where the operator is recast in the form
\[
|A\rangle\rangle=\sum_{jk}a_{jk}|jk\rangle\rangle,
\]
where $|jk\rangle\rangle \equiv |jk\rangle \langle jk|$. The thermofield representation simply encodes these coefficients as the {\it amplitudes} of a quantum state. Since we need $d \times d$ entries to define an operator, we define an ``ancillary'' space, which is an exact duplicate of our original Hilbert space\[
\mathcal{H} \rightarrow \mathcal{H} \otimes \mathcal{H}.
\]
For each state $|x\rangle$ in $\mathcal{H}$, we define a ``tilde'' state $|\tilde x\rangle$ in the ancillary space, which is an exact copy. We can now define a quantum wave-function as:
\begin{equation}
|\psi_A\rangle = \sum_{jk}a_{jk}|j\rangle|\tilde k \rangle.
\label{thermofield}
\end{equation}
Same as in Liouville representation, thermo-field dynamics establishes a framework to work with operators. The advantage of this formalism is that operators are now wave-functions, and super-operators become conventional quantum mechanical operators acting on these wave-functions.  
Therefore, we can apply all our zero-temperature many-body machinery with the price of working in a larger Hilbert space.

These ideas can readily be generalized to an operator such as the thermal density matrix $\rho=\exp(-\beta H)$, where $\beta=1/T$ is the inverse temperature, and $H$ is some Hamiltonian under consideration.
We would like to calculate the thermal average of the operator $A$:
\begin{equation}
\langle A \rangle = Z(\beta)^{-1} \mathrm{Tr}(\rho A); \,\,\, Z(\beta) = \mathrm{Tr}(\rho).
\label{average}
\end{equation}
The thermo-field representation $\psi(\beta)$ of the thermal density matrix allows us to write this thermal average as a conventional expectation value of an operator in a pure quantum state:
\begin{equation}
\langle A \rangle_{\beta} = \frac {\langle \psi(\beta) | A | \psi(\beta) \rangle}{\langle \psi(\beta) | \psi(\beta) \rangle}.
\label{average}
\end{equation}
Here $| \psi(\beta) \rangle $ can is writen in the basis of energy eigenstates of the system under study, $\{n\}$, as
\begin{equation}
| \psi(\beta) \rangle = e^{-\beta H/2}| \psi(0) \rangle = \sum_n
e^{-\beta E_n/2} |n \tilde{n}\rangle \label{thermoa}
\end{equation}
where $|\psi(0)\rangle=\sum_n{|n\tilde{n}\rangle}$ is our thermal vacuum.

For illustration let us consider a two-level system. We can write the thermal density matrix as
\[
\rho = \rho_{00}|0\rangle \langle0| + \rho_{01}|0\rangle \langle1| + \rho_{10}|1\rangle \langle0| + \rho_{11}|1\rangle \langle1|
\]
or
\[
|\psi\rangle = \rho_{00}|0\rangle |\tilde0\rangle + \rho_{10}|1\rangle |\tilde0\rangle + \rho_{01}|0\rangle |\tilde1\rangle + \rho_{11}|1\rangle |\tilde1\rangle.
\]
In particular, at infinite temperature, this becomes
\[
\rho_0 = \frac{1}{2} |0\rangle \langle0| + \frac{1}{2} |1\rangle \langle 1|
\]
or 
\[
|\psi_0 \rangle = \frac{1}{2} |0\rangle|\tilde 0 \rangle + \frac{1}{2} |1\rangle|\tilde 1 \rangle
\]
In general we find it convenient to perform a particle-hole transformation on the ancillas. Our state then becomes
\[
|\psi_0 \rangle = \frac{1}{2} |0\rangle|\tilde 1 \rangle + \frac{1}{2} |1\rangle|\tilde 0 \rangle.
\]
For the case of of $S-1/2$ spins, for instance, this state can be written as $|\psi_0\rangle=1/2|\uparrow,\tilde{\downarrow}\rangle + 1/2|\downarrow,\tilde{\uparrow}\rangle$ \cite{Feiguin2005a}. It is important to notice that the sign, or whether the maximally entangled state under consideration is a singlet or a triplet, does not change the results.

At $\beta=0$, the state $|\psi_0\rangle$ is the maximally entangled state between
the real system and the fictitious system.  
 It is natural to work in an occupation number representation
where the state of each site $i$ takes on a
definite value $n_i$. For a many-body system, one finds
\begin{equation}
| \psi(\beta=0) \rangle = \prod_i \sum_{n_i} |n_i \tilde{n}_i\rangle = \prod_i
|\psi_{0i}\rangle \label{site}, \,
\end{equation}
defining the maximally entangled state $|\psi_{0i}\rangle$ of site $i$ with
its ``ancilla'', the local degree of freedom in the auxiliary system. Notice that the maximally entangled state is not uniquely defined (as seen before, a triplet and a singlet are both maximally entangled states, for instance).

The state of the system at an arbitrary temperature $\beta$ is obtained by
evolving the maximally mixed state in imaginary time, Eq.~(\ref{thermoa}) with $\beta=0$, using
the Hamiltonian acting on the real degrees of freedom. The ancillas do not have
any interactions controlling their dynamics. They evolve only by their
entanglement with the physical spins, effectively acting as a thermal bath. This 
is the basis of the finite-temperature DMRG method \cite{Feiguin2005a} (We remark that it is also possible to use a time dependent evolution to generate the purified state, see, e.g. \cite{Mann1989197}). Notice
that at zero temperature, the site and the ancilla are totally disentangled,
while at finite temperature there is always a finite degree of entanglement that 
only depends on the dynamics of the system. This process is also referred-to as quantum purification.

We now show how to represent a Hamiltonian whose ground state realizes the thermofield dynamics state. Let ${\cal H}$ we be the original Hilbert space, which, we assumed to be factorized as $\otimes {\cal H}_i$ where ${\cal H}_i$ are local degrees of freedom such as spins, with a basis
$|{\sigma_l}^{(j)}\rangle$.

We now define the Hamiltonian on the doubled system ${\cal H}\otimes
{\cal H}$ as
\begin{eqnarray}
H_0=\Sigma _lH_{l,0 }\text{         };\text{    }H_{l,0}=\left(1-\left|\psi _{0,l}\right\rangle \left\langle \psi _{0,l}\right|\right) 
\end{eqnarray}
{where }
\begin{eqnarray} \left|\psi _{0,l}\right\rangle =\Sigma _{\sigma }|\sigma \rangle _l\otimes |\sigma \rangle _l \end{eqnarray}
projects on the maximally entangled state of site $l$.  $H_{0}$ is associated with taking the infinite temperature limit $\beta\rightarrow 0$, and it's  ground state  { is simply:}
\begin{eqnarray}
\psi _0=\otimes _l\left|\psi _{0,l}\right\rangle
\end{eqnarray}
{To access other temperatures,  associated with a finite inverse temperature $\beta$ we write a new Hamiltonian  as follows:}
\begin{eqnarray}
H_{\beta }=\left(e^{-\beta  \frac{H}{2}}\otimes I\right)H_0\left(e^{\beta  \frac{H}{2}}\otimes I\right)=\Sigma _lH_{l,\beta }\label{Thermal transformation}
\end{eqnarray}
\text{where }
\begin{eqnarray}
H_{l,\beta }=1-\left(e^{-\beta  \frac{H}{2}}\otimes I\right)\left|\psi _{0,l}\right\rangle \left\langle \psi _{0,l}\right|\left(e^{\beta  \frac{H}{2}}\otimes I\right)
\end{eqnarray}

Note that since $H_{\beta}$ is related to $H_{0}$ by a similarity
transformation, the two operators share eigenvalues, in this case 
non-negative integers. In particular, this common spectrum has the important
consequence that the new Hamiltonian $H_{\beta}$ is gapped. Moreover, since the
initial local terms of $H'$ commute, so will the transformed
elements. 
While the ground states of each of these local Hamiltonians
have an extensive degeneracy, we are looking for the unique ground state of the sum. 

The Hamiltonian $H_{\beta}$ is a sum of (non Hermitian)
constraints, which are all satisfied by the ground state.  Thus, one has to be careful about making statements and calculations with it. 
One can have a Hermitian version of it by using instead:
\begin{eqnarray}
H_{\beta }^{\text{Hermitian}}=\Sigma _lH_{l,\beta }{}^+H_{l,\beta }
\end{eqnarray}
{Since all the terms are positive, and }
$H_{l,\beta }$
each annihilate 
$\psi _{\beta }$
we are assured that 
$\psi _{\beta }$ { is also the ground state of }
$H_{\beta }^{\text{Hermitian}}$. Note, however, that $H_{\beta }^{\text{Hermitian}}$ will, in general, have a completely different spectrum than $H_{\beta }$.


\begin{figure}[hb]
\includegraphics*[width=3.5in]{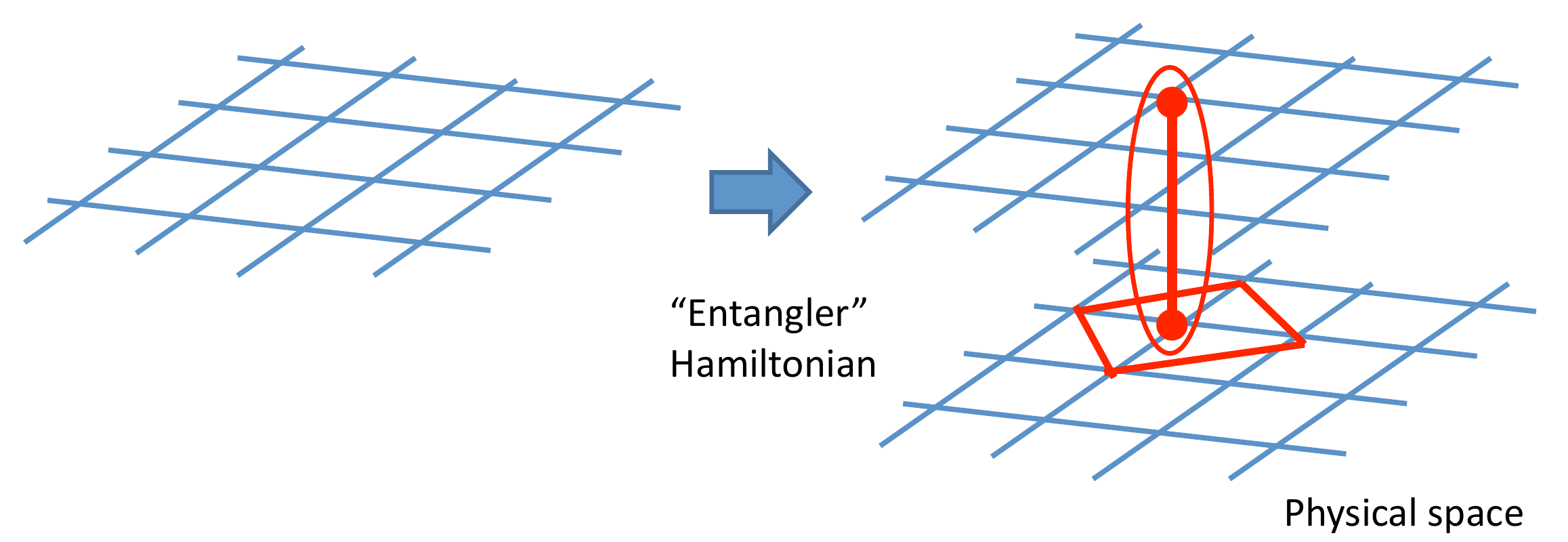}
\caption{Thermal Hamiltonian representation for the 2D Ising model. The Hilbert space is doubled, and the Hamiltonian acts on the resulting bi-layer system. Each term involves operators acting on four neighbors of a site and the ``entangler'' connects the layers, introducing the temperature in the form of quantum fluctuations.
} 
\label{fig1}
\end{figure}
\paragraph{Classical and quantum criticality.}  A particularly convenient class of Hamiltonians for which we can construct explicitly the thermal Hamiltonian $H_{\beta}$ are classical Hamiltonians. In such Hamiltonians, since all terms commute, the transformation (\ref{Thermal transformation}) can be determined locally at any temperature $\beta$. 
This approach allows us to explore the relation between classical and quantum criticality. For example, it is well known (see e.g. \cite{sondhi1997continuous}) that critical exponents in the vicinity of quantum critical points in $d$ dimensions are characterized by finite size corrections to corresponding $d$ dimensional  classical theories (in contrast with the usual relation of $d$ quantum mechanical systems with $d+1$ classical systems). 

The study of quantum critical models with amplitudes given by Boltzman weights of classical systems has been very fruitful in studying quantum criticality and topological quantum states such as loop gases a prime example of which is  Kitaev's toric code \cite{kitaev2003fault}. Additional prominent examples are the quantum dimer model \cite{rokhsar1988superconductivity}, and the quantum Lifshitz model \cite{ardonne2004topological}.  Quantum critical points associated with such 2d classical models are now referred to as conformal quantum critical points.
There is also a growing interest in constructing explicit Hamiltonians which realize models of this type \cite{cano2010spin,Laumann2010}.
The relation of quantum models of this type to classical stochastic dynamics is highlighted Refs.\cite{Castelnovo2010,isakov2011dynamics}, where the classical state is generated by  a classical master equation which essentially implements thermofield dynamics. In \cite{verstraete2006criticality,somma2007quantum} the construction of quantum models associated with classical models have been highlighted by generating appropriate ``classical thermal'' coherent states. These states are pure states and do not require doubling the system as done here. On the other hand, these states can only be used directly to compute observables which are diagonal in the energy basis.

While the thermal Hamiltonians we generate using our methods may be complicated, the advantage of the approach taken here is that it is an explicit construction, applicable as a general framework for any classical model of interest. As an illustration, consider the Ising model:
\begin{eqnarray}
H= \Sigma J_{i,j}\sigma _i^z \sigma _j^z +h \sigma _i^z
\end{eqnarray}
for a spin 1/2 system we can take our $\beta=0$ Hamiltonian as the singlet projection:
\begin{eqnarray}
H_0=\Sigma_i (\sigma _i\cdot \tilde{\sigma }_i+3)
\end{eqnarray}
application of the thermal transformation yields:
\begin{eqnarray}
H_{l,\beta }=3+\sigma ^z_{l}\tilde{\sigma }^z_{l}+\sum_{\mu,\nu\in\{x,y\}}K_{\mu  \nu }\sigma _{\mu ,l}\tilde{\sigma }_{\nu ,l}
\end{eqnarray}
where $K_{\mu  \nu }$ is an operator which depends on the configuration of neighboring spins in the $\sigma_z$ basis.

In the particular, for the Ising model with nearest neighbors on a square lattice, we find:
\begin{eqnarray}&
K_{yy}=K_{x x}=(Q v^4+\frac{1}{2} R^2 u^2 v^2+u^2 (1-v^2))\cosh(\beta h)+\nn & u v  R \left(Q  v^2+ u^2\right)\sinh(\beta h)\\ &
K_{x y}=-K_{x y}=i u v  R\left(Q  v^2+ u^2\right)\cosh(\beta h)+\nn & i  \left(Q v^4+\frac{1}{2} R^2 u^2 v^2+u^2 (1-v^2)\right)\sinh(\beta h)
\end{eqnarray}
where $u=\cosh (\beta  J), v=\sinh (\beta  J)$ and $Q,R$ are plaquette and sum operators:
\begin{eqnarray}
R=\sum_{i:\langle ij\rangle}\sigma ^z_{l}~~~;~~~Q=\prod_{i:\langle ij\rangle}\sigma ^z_{l}
\end{eqnarray}
alternatively, this Hamiltonian may be written as:
\begin{eqnarray*}
H_{l,\beta }=3+\sigma ^z_{l}\tilde{\sigma }^z_{l}+{K_{xx}-K_{yy}\over 2}\sigma^+_{l}\tilde{\sigma }^-_{l}+{K_{xx}-K_{yy}\over 2}\sigma^-_{l}\tilde{\sigma }^+_{l}.
\end{eqnarray*}
We have thus a quantum Hamiltonian whose ground state is given explicitly by the TFD representation with simple Boltzmann weights.

\paragraph{ Realization using DMRG.}
One way to explore and analyze the expansion of the thermal Hamiltonian in a quantum many-body system is by directly studying an example. As a proof of concept, we chose the isotropic Heisenberg chain, and compared the energies obtained from the ground state of the thermal Hamiltonian, with the exact results obtained by other means, such a temperature DMRG\cite{Feiguin2005a} and exact diagonalization. Instead of evolving in imaginary time, in this case we just diagonalize the thermal Hamiltonian and obtain its ground-state. Notice two important factors: the Hamiltonian is non-Hermitian, and the Hilbert space has twice the number of spins. In order to diagonalize a non-Hermitian Hamiltonian, we use the techniques learned from transfer matrix renormalization group (TMRG) \cite{Nishino1995,Bursill1996,Wang1997}, and apply a modified power method instead of the usual Lanczos/Davidson approach.
In the non-Hermitian version of DMRG, one has to obtain the right and left eigenvectors of the Hamiltonian $|\psi_L\rangle,|\psi_R\rangle$, and build the reduced density matrix for the truncation. For simplicity, we considered a symmetric form $\rho=1/2|\psi_L\rangle\langle\psi_L|+1/2|\psi_R\rangle \langle\psi_R|$ \cite{Nishino1999,Enss2001}. This approach works well in practice but breaks down at low temperatures yielding complex energies. For that reason, and since our main goal is to understand the behavior of the thermal Hamiltonian, at low temperatures we performed exact diagonalization on smaller systems, and avoided the truncation (It is very likely that more sophisticated truncation schemes might be able to solve this problem \cite{Huang2012}).
As we lower the temperature, and the correlation length grows, we need to consider higher order expansions. In practice, in order to perform the calculation, we just apply multiple products of the Hamiltonian $H$ and $H_0$ (we do not have to calculate the thermal Hamiltonian explicitly!).
We have found that a first order expansion yields very accurate results at high temperature $T>J$. This is consistent with a small correlation length, due to the thermal fluctuations. It is reasonable then to expect that the first order, which contains only local terms, would be a good approximation. At lower temperatures the convergence is slow, and resembles the behavior of a high-temperature series expansion (HTE)\cite{HTE}. Although there is not a straightforward connection between both techniques, one could assume that the algorithm is effectively implementing quantum parallellism with the ancillas, and summing over all possible expansion diagrams in the HTE.
As a curious fact, notice that the non-Hermitian thermal Hamiltonian yields the thermal density matrices of both the ferromagnetic and antiferromagnetic Heisenberg chain, in the form of the left and right eigenvectors.

\begin{figure}[ht]
  \centering
  \includegraphics[width=0.45\textwidth]{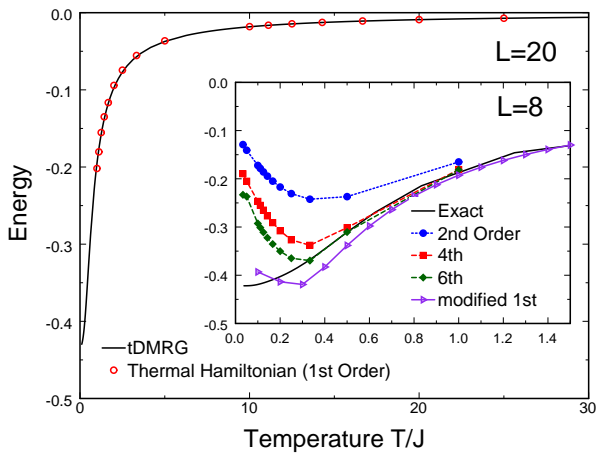}
  \caption{Proof of concept numerical results for a small Heisenberg chain. We show the energy as a function of temperature for the non-Hermitian thermal Hamiltonian. At high temperature $T>J$, a first order expansion sufices and yields excellent results. At lower temperature, the correlation length grows, and the accuracy improves as we increase the order of the expansion. The modified 1st order expansion was obtained with the Hamiltonian (\ref{alternative thermal}). Numerical details are explained in the text.
}
  \label{fig:dmrg}
\end{figure}

\paragraph {Properties of the thermal Hamiltonians.} The class of Hamiltonians we obtain have fascinating properties. For example, the non Hermitian Hamiltonians are sums of commuting terms with integer spectrum which is bounded below by zero. However, this does not mean that the actual ground state described is gapped, and has exponentially decaying correlations. Indeed, as we have seen, it is easy to construct states which are critical, using this approach. 
Essentially, the relation between gap and correlation decay holds only for Hermitian Hamiltonians. In particular, since states associated with different eigenvalues may be non orthogonal.

Our Hamiltonian belongs to the class of frustration free Hamiltonians. Such Hamiltonians are of great interest due to their simplicity and appearance in many quantum systems, prominent examples are the (ferromagnetic) Heisenberg chain,  AKLT\cite{affleck1987rigorous}, and the toric code\cite{kitaev2003fault}. Efficient methods for solving this class of Hamiltonians are available \cite{feiguin2013exact}. We note that frustration free Hamiltonians, may still yield highly non trivial states even in 1D, see \cite{bravyi2012criticality} for an example of a spin-1 chain with a highly entangled ground state. We also point out that the ``frustration free'' denomination is usually applied to local Hamiltonians, which may or may not be the case in our formulation. 

We emphasize that thermal Hamiltonians are not unique, as is the case for any frustration free Hamiltonian. One may easily construct other Hamiltonians with the same ground state. 
For example, choose any set of positive operators $A_l$. Then the Hamitlonian:
\begin{eqnarray}
H_{\beta , A}=\Sigma_l H_{l,\beta } A_l H_{l,\beta }, \label{alternative thermal}
\end{eqnarray}
will still have a  $\psi_{\beta}$ as a ground state with eigenvalue $0$.  

One may make use of this freedom via a wise choice of $A_l$
to improve the convergence properties of numerical computations. In fact, note that $H_{\beta , A}=\Sigma_l A_l H_{l,\beta }$, will also have $\psi_{\beta}$ as a zero energy eigenvector (but the spectrum will not be necessarily non-negative). A particularly simple choice is: $A_l=H_{0,l}$. In Fig.~\ref{fig:dmrg}, we compare the hamiltonian $H_{\beta}$ of the Heisenberg spin chain expanded to first order in $\beta$ with $A_l=H_{0,l}$ with expansions of $H_{\beta}$ in $\beta$ to higher orders, and see that adding $A_l$ substantially improves convergence at low temperatures. Unfortunately, it is not clear at the moment if there is a systematic way of finding what is the optimal $A_l$ one should in a given problem.

\paragraph{Discussion}
In this work we have demonstrated an elementary approach to obtaining TFD states as ground states of concrete Hamiltonians. The obvious limitation of the proposed method is the long range nature of the Hamiltonians at low temperatures. However, as we have seen, for ``classical'' models and simple quantum models (such as the toric code) the resulting Hamiltonians are short range at any temperature. Moreover, for any model, the thermal Hamiltonian at $T=\infty$ is local, and certainly, at $T=0$ we can represent the ground state as that of the original, local, Hamiltonian. A natural question that arises is: what are the conditions on the existence of a short-range Hamiltonian realizing the TFD picture for any temperature $T$, and is there an efficient way of finding such a Hamiltonian? In fact, there are indications that such a Hamiltonian might exist for many systems of interest \cite{Poilblanc2010,Feiguin2011,Soltanieh-ha2012,Fradkin2013}. It is easy to see that the von Neumann entanglement entropy between the physical subsystem and the ancillas is indeed equal to the thermal Gibbs entropy $S=\beta E$, a correspondence that was observed to hold to some extent in the ground state of spin ladders. Another interesting aspect of the non-Hermitian thermal Hamiltonians is that their spectrum remains {\it gapped} across a quantum critical point, but the eigenstates are non-orthogonal, and the ground-state possesses algebraically decaying correlations. We believe this indicates that this new family of models may hide interesting features 
and may require further exploration.

\paragraph{Acknowledgments}
AEF thanks T. Nishino for assistance with the non-Hermitian diagonalization. We acknowledge useful discussions with Paul Fendley.
AEF and IK are grateful to NSF for funding under grants DMR-1339564 and DMR-0956053, respectively.\bibliographystyle{apsrev}
\bibliography{TFD.bib}

\end{document}